\documentclass[preprint,preprintnumbers]{revtex4}

\usepackage{graphicx}
\usepackage{dcolumn}
\usepackage{bm}
\usepackage{epsfig}
\usepackage{epstopdf}
\usepackage{grffile}
\usepackage{color}
\usepackage{colordvi}
\usepackage{array}
\usepackage{tikz} 
\usepackage{tikz-feynman}
\usepackage{amsmath}
\usepackage{amssymb}
\usepackage{rotating}
\usepackage{lscape}
\usepackage{float}
\usepackage{footnote}
\usepackage{wasysym}
\usepackage{gensymb}
\usepackage{subcaption}
 \makesavenoteenv{environmentname}

\usepackage{hyperref}
\usepackage{ragged2e}
\makeatletter
\long\def\@makecaption#1#2{%
	\vskip\abovecaptionskip
	\sbox\@tempboxa{#1: #2}%
	\ifdim \wd\@tempboxa >\hsize
	#1: \justifying #2\par
	\else
	\global \@minipagefalse
	\hb@xt@\hsize{\hfil\box\@tempboxa\hfil}%
	\fi
	\vskip\belowcaptionskip}
\makeatother
\begin{document}


\title{Freeze-in Asymmetric Dark Matter with a Dual-Role Scalar: CP Violation and Symmetric-Component Depletion}

\author{S. Peyman Zakeri}
\affiliation{Faculty of Physics, Yazd University, P.~O.~Box 89195--741, Yazd, Iran}

\date{\today}

\begin{abstract}
	We present a minimal, predictive scenario for asymmetric fermionic dark matter (DM) generated through freeze-in. The standard model (SM) is extended by a Dirac fermion $\psi$ that serves as the DM candidate, an auxiliary Dirac fermion $\chi$, a real scalar mediator $\phi$, and a heavier real scalar $S$, all stabilized by a $Z_2 \times Z_2'$ discrete symmetry. The DM asymmetry is generated through the CP-violating out-of-equilibrium decays of the mediator $\phi$, while the heavier scalar $S$ plays a dual role: it generates the physical CP-violating phase required for the asymmetry through its interference with $\phi$, and it mediates the annihilation of the symmetric component of the DM population. We solve the coupled Boltzmann equations for the asymmetric yield $Y_-$ and the total yield $Y_\Sigma$, and identify the scaling relations $Y_-^\infty \propto \mu^2 \varepsilon / m_\phi$ and $\Omega_{\rm DM} h^2 \propto m_\psi \, \mu^2 \varepsilon / m_\phi$, which we verify numerically. The observed relic abundance $\Omega_{\rm DM} h^2 \simeq 0.12$ is naturally reproduced for a wide range of parameters; in particular, for $\varepsilon = 10^{-6}$ and $m_\phi = 200$~GeV, the correct relic density is obtained for $\mu \approx 2.7 \times 10^{-9}$~GeV and $m_\psi \approx 40$~GeV. We further confront the model with current experimental constraints, including direct detection, indirect detection, collider searches, and DM self-interactions, and show that the predicted signals lie far below the sensitivity of current experiments. This scenario provides a self-consistent and testable realization of asymmetric fermionic DM within the freeze-in framework.
\end{abstract}

\maketitle


\section{Introduction}\label{sec:int} 
Understanding the nature of DM remains one of the most compelling challenges at the interface of particle physics, astrophysics, and cosmology. Although the gravitational imprint of DM has been detected across an enormous range of scales—from the dynamics of dwarf galaxies to the acoustic peaks of the cosmic microwave background—its microscopic properties are still entirely unknown~\cite{Bertone:2005,Feng:2010}. The standard cosmological model successfully describes the large-scale structure of the Universe only if DM is cold, non-baryonic, and stable on cosmological timescales, yet none of the known particles of the SM can fulfill this role.

Among the numerous DM candidates proposed in the literature, the weakly interacting massive particle (WIMP) has traditionally received the most attention, owing to its ability to reproduce the observed relic density via the thermal freeze-out mechanism~\cite{Zeldovich:1965,Chiu:1966}. In this scenario, DM particles are kept in thermal equilibrium with the SM bath by rapid annihilation processes until the Hubble expansion rate drops below the interaction rate, at which point they decouple. Despite extensive experimental efforts over the past decades—ranging from direct detection experiments such as LUX-ZEPLIN and XENONnT to collider searches at the LHC—no definitive evidence for WIMP DM has emerged~\cite{LZ:2023,XENONnT:2023}. The absence of any signal, together with the lack of new physics signatures at the TeV scale, has prompted a growing interest in alternative production mechanisms.

A radically different proposal, known as the freeze-in mechanism, has emerged as a compelling framework for generating the observed DM abundance without invoking thermal equilibrium~\cite{Hall:2009,Bernal:2017}. In this picture, the dark sector interacts so weakly with the visible sector that its population builds up incrementally through rare production processes, such as the decay or annihilation of particles in the primordial plasma. In this scenario, often referred to as feebly interacting massive particle (FIMP), the final relic density is set by the strength of the feeble couplings rather than by the details of thermal decoupling, offering a natural setting for DM candidates that would otherwise remain invisible to conventional searches. This idea has been realized in numerous extensions of the SM, involving scalar, fermionic, or vector DM fields~\cite{Zakeri:2018,Ayazi:2016,Bernal:2016}. Comprehensive accounts of the freeze-in mechanism and its phenomenological implications can be found in Refs.~\cite{Bernal:2017,Belanger:2024}.

An alternative framework that has attracted considerable attention is asymmetric DM (ADM), in which the present-day DM density is governed by a primordial particle–antiparticle asymmetry in the dark sector rather than by thermal freeze-out~\cite{Petraki:2013,Zurek:2013,Kaplan:2009}. The principal motivation for this scenario stems from the observation that the energy densities of DM and baryons are remarkably close, with $\rho_{\rm DM} \simeq 5\rho_B$~\cite{Petraki:2013,Zurek:2013}. In the ADM picture, this numerical proximity finds a natural explanation if the dark and visible asymmetries share a common origin or are generated by analogous mechanisms. A crucial requirement for such models is that the symmetric component of the DM population be efficiently depleted, so that the final abundance is dominated by the asymmetric population~\cite{Graesser:2011}.

The asymmetric freeze-in (AFI) framework combines the essential features of both ADM and freeze-in: the DM abundance is generated out of equilibrium while simultaneously carrying a particle–antiparticle asymmetry~\cite{Hall:2010,Unwin:2014}. In this scenario, the asymmetry is typically produced through the CP-violating out-of-equilibrium decays of a heavy mediator that remains in thermal equilibrium with the visible sector but decoupled from the dark sector. This requirement is crucial, as a mediator in thermal equilibrium with the dark sector can significantly suppress the generated asymmetry~\cite{Hook:2011}. The AFI mechanism has been explored in various contexts with different field content (scalar, fermion, ...)~\cite{Unwin:2014,Tsao:2017}, and also in models with more than one DM candidate~\cite{Zakeri:2026,Beneito:2023}. In the present work, we focus on a fermionic DM candidate and explore a scenario in which the asymmetry is generated through the out-of-equilibrium decay of a scalar mediator, while the symmetric component is efficiently depleted by annihilations mediated by a heavier scalar field.

This paper is organized as follows. In Section~\ref{sec:model}, we present the model setup, including the particle content, the discrete symmetries, and the Lagrangian. In Section~\ref{sec:boltz}, we derive the Boltzmann equations, compute the asymmetric and total yields, and determine the relic abundance. Section~\ref{sec:aspect} is devoted to the phenomenological implications of the model, namely direct detection, indirect detection, collider constraints, and self-interactions. Finally, we summarize our results in Section~\ref{sec:conclud}.

\section{The Model}\label{sec:model}
The dark sector that we study here contains four new fields: a Dirac fermion $\psi$, which is the DM candidate, a second Dirac fermion $\chi$ that plays an auxiliary role, a real scalar $\phi$ that acts as the mediator, and a heavier real scalar $S$. All new fields are singlets under the SM gauge group and interact with the visible sector solely through the Higgs portal. The model respects a $Z_2 \times Z_2'$ discrete symmetry, under which $\psi$ and $\chi$ are charged as $(-,+)$, $\phi$ and $S$ as $(+,-)$, and all SM fields are neutral. This symmetry ensures the stability of the DM candidate $\psi$ and forbids undesired operators such as $\phi^3$, $S^3$, and $\phi H^\dagger H$. The masses of the new fields are ordered as $m_\psi < m_\chi < m_\phi < m_S$, so that $\psi$ cannot decay into any other new particle.

The Lagrangian of the dark sector is given by
\begin{equation}
	\begin{aligned}
		\mathcal{L}_{\rm dark} \supset \; &
		\frac{1}{2}\partial_\mu \phi\, \partial^\mu \phi
		+ \frac{1}{2}\partial_\mu S\, \partial^\mu S
		+ i \bar \psi \gamma^\mu \partial_\mu \psi
		+ i \bar \chi \gamma^\mu \partial_\mu \chi \\
		& - \frac{1}{2} m_\phi^2 \phi^2
		- \frac{1}{2} m_S^2 S^2
		- m_\psi \bar \psi \psi
		- m_\chi \bar \chi \chi \\
		& - \left(
		\mu\, \phi\, \bar \chi \psi
		+ y_S\, S\, \bar \chi \psi
		+ \lambda_S\, \phi\, S^2
		+ y_\psi\, S\, \bar \psi \psi
		+ \text{h.c.}
		\right) \\
		& - \lambda_{SH}\, S\, H^\dagger H
		- \lambda_{\phi H}\, \phi^2\, H^\dagger H
		- V(H) ,
	\end{aligned}
	\label{eq:Lagrangian}
\end{equation}
where $H$ is the SM Higgs doublet and $V(H)$ is the SM Higgs potential,
\begin{equation}
	V(H) =
	- \mu_H^2 H^\dagger H
	+ \lambda_H (H^\dagger H)^2 .
	\label{eq:potential}
\end{equation}
The trilinear coupling $\mu\,\phi\bar\chi\psi$ governs the out-of-equilibrium decay of the mediator and is responsible for the freeze-in production of the dark sector. The couplings $y_S\,S\bar\chi\psi$ and $\lambda_S\,\phi S^2$ generate a physical CP-violating phase through the interference between the tree-level and loop-level decay amplitudes of $\phi$. The coupling $y_\psi\,S\bar\psi\psi$ controls the annihilation of the symmetric component of the DM population, while its smallness ensures that the late-time decay $S \to \psi\bar\psi$ remains negligible. Moreover, the Higgs-portal couplings $\lambda_{\phi H}$ and $\lambda_{SH}$ maintain thermal contact between the dark sector and the SM plasma.

The DM candidate $\psi$ is rendered stable by the $Z_2$ symmetry, under which it is the lightest charged particle. This symmetry also forbids any loop-induced decay of $\psi$. The auxiliary fermion $\chi$ plays a twofold role: it appears as a final state in the decay $\phi \to \chi \psi$, which is the primary source of the DM population, and it mediates the CP-violating interference through the coupling $y_S\,S\bar\chi\psi$. The scalar mediator $\phi$ remains in thermal equilibrium with the SM plasma via the Higgs portal, while the dark fermions $\psi$ and $\chi$ never thermalize owing to the feeble coupling $\mu$, thereby realizing the freeze-in regime. The heavier scalar $S$ enters the dark dynamics in two distinct ways. At the loop level, its exchange interferes with the tree-level $\phi$ amplitude and supplies the physical phase that makes the CP asymmetry $\varepsilon$ non-zero. At the same time, the $s$-channel process $\psi \bar{\psi} \to S \to \mathrm{SM}$ removes the symmetric part of the dark population through the coupling $y_{\psi} S \bar{\psi} \psi$; this channel is efficient in the high-temperature regime but becomes Boltzmann-suppressed once the Universe cools well below $m_{S}$.

The model is characterized by four new masses and six new couplings. The DM mass is taken in the range $m_\psi \sim \mathcal{O}(1\text{--}100)$~GeV, while the auxiliary fermion and the two scalars satisfy $m_\chi \gtrsim m_\psi$, $m_\phi > 2m_\psi$, and $m_S > m_\phi$. The trilinear coupling $\mu$ is chosen sufficiently small to satisfy the freeze-in condition $\Gamma_\phi < H(T \sim m_\phi)$~\cite{Hall:2009}, ensuring that the mediator never thermalizes with the dark fermions. The couplings $y_S$ and $\lambda_S$ control the size of the CP-violating parameter $\varepsilon$ and must be large enough to generate a viable asymmetry, while remaining perturbative~\cite{Hook:2011}. The coupling $y_\psi$ is constrained by two competing requirements: it must be large enough to allow efficient annihilation of the symmetric component at high temperatures~\cite{Graesser:2011}, yet small enough to suppress the late-time decay $S \to \psi\bar\psi$. Finally, the Higgs-portal couplings $\lambda_{\phi H}$ and $\lambda_{SH}$ are taken large enough to keep the scalar mediators in thermal equilibrium with the SM plasma, while the feeble coupling $\mu$ ensures the out-of-equilibrium production of the dark sector. These conditions define a well-defined and predictive region of parameter space, which will be explored numerically in the following sections.

\section{Relic abundance}\label{sec:boltz} 

\subsection{Boltzmann equation}\label{sec3.1}
The DM abundance in this framework is set by the freeze-in mechanism, with the dark sector populated through the decays of the scalar mediator $\phi$ while this decay occurs out of equilibrium with the visible plasma~\cite{Hall:2009}. In this regime, $\phi$ is kept in thermal contact with the SM bath through the Higgs-portal coupling $\lambda_{\phi H}$, whereas $\psi$ and $\chi$ never reach thermal equilibrium. Consequently, the dark sector is populated gradually through rare decays $\phi \to \chi \psi$, and inverse processes such as $\chi \psi \to \phi$ can be safely neglected throughout the cosmological evolution~\cite{Hall:2009}. In this regime, the evolution of the dark-sector abundances is described by a set of coupled Boltzmann equations for the asymmetric yield $Y_-$ and the total yield $Y_\Sigma$, where the latter accounts for the symmetric component of the DM population and its subsequent annihilation.

The generation of the dark-sector asymmetry is parametrized by the CP-violating parameter $\varepsilon$, defined as
\begin{equation}
	\varepsilon = \frac{\Gamma(\phi \to \chi \psi) - \Gamma(\phi \to \bar\chi \bar\psi)}{\Gamma_\phi},
	\label{eq:epsilon}
\end{equation}
where $\Gamma_\phi$ is the total decay width of the scalar mediator. A non-zero value of $\varepsilon$ arises from the interference between the tree-level decay amplitude and the one-loop corrections involving the heavier scalar $S$, with the couplings $y_S$ and $\lambda_S$ providing the necessary physical phases~\cite{Hook:2011}. Since the model contains a single fermionic decay channel, the CP asymmetry is generated entirely at the loop level, in contrast to scenarios where multiple tree-level channels interfere.

The asymmetry yield is defined as
\begin{equation}
	Y_- \equiv Y_\psi - Y_{\bar\psi}
	=
	\frac{n_\psi - n_{\bar\psi}}{s},
	\label{eq:yield_asym}
\end{equation}
where $s$ denotes the entropy density of the Universe. Assuming that the scalar mediator remains in thermal equilibrium throughout the production epoch, the evolution of the asymmetric number density $n_- = n_\psi - n_{\bar\psi}$ is governed by the Boltzmann equation~\cite{Hook:2011}
\begin{equation}
	\frac{dn_-}{dt} + 3H n_- = \varepsilon \, \Gamma_\phi \, n_\phi^{\rm eq},
	\label{eq:boltz_asym}
\end{equation}
where $n_\phi^{\rm eq}$ is the equilibrium number density of the mediator. Using the relation $d/dt = -HT\,d/dT$ and the entropy density
\begin{equation}
	s(T) = \frac{2\pi^2}{45} g_* T^3,
	\label{eq:entropy}
\end{equation}
with $g_*$ the effective number of relativistic degrees of freedom, the Boltzmann equation can be rewritten in terms of the yield as
\begin{equation}
	\frac{dY_-}{dT}
	=
	- \frac{\varepsilon \, \Gamma_\phi \, n_\phi^{\rm eq}(T)}
	{s(T)\, H(T)\, T}.
	\label{eq:boltz_yield}
\end{equation}
The equilibrium number density of the scalar mediator is given by~\cite{Gondolo:1991}
\begin{equation}
n_\phi^{\rm eq}(T) = \frac{g_\phi \, m_\phi^2 T}{2\pi^2} K_2\!\left( \frac{m_\phi}{T} \right),
\end{equation}
where $g_\phi = 1$ is the number of internal degrees of freedom of the real scalar mediator. Furthermore, the Hubble expansion rate during radiation domination reads~\cite{Kolb:1990}
\begin{equation}
	H(T) = 1.66 \, \sqrt{g_*} \, \frac{T^2}{M_{\rm Pl}},
	\label{eq:hubble}
\end{equation}
with $M_{\rm Pl}$ the Planck mass.

The total yield $Y_\Sigma$, which includes both the asymmetric and symmetric components of the DM population, is defined as
\begin{equation}
	Y_\Sigma \equiv Y_\psi + Y_{\bar\psi}
	=
	\frac{n_\psi + n_{\bar\psi}}{s},
	\label{eq:yield_sym}
\end{equation}
whose evolution is governed by the Boltzmann equation~\cite{Graesser:2011}
\begin{equation}
	\frac{dn_\Sigma}{dt} + 3H n_\Sigma
	=
	2\,\Gamma_\phi \, n_\phi^{\rm eq}
	-
	2\,\langle \sigma v \rangle \left( n_\psi n_{\bar\psi} - n_\psi^{\rm eq} n_{\bar\psi}^{\rm eq} \right),
	\label{eq:boltz_sym}
\end{equation}
where $n_\Sigma = n_\psi + n_{\bar\psi}$ and $\langle \sigma v \rangle$ is the thermally averaged annihilation cross section~\cite{Gondolo:1991} of the process $\psi\bar\psi \to S \to \text{SM}$. The first term on the right-hand side accounts for the production of DM particles and antiparticles through mediator decays, while the second term describes the annihilation of the symmetric component. In terms of the yield, the Boltzmann equation takes the form
\begin{equation}
	\frac{dY_\Sigma}{dT}
	=
	-\frac{2\,\Gamma_\phi \, n_\phi^{\rm eq}(T)}{s(T)\, H(T)\, T}
	+
	\frac{2\,\langle \sigma v \rangle}{s(T)\, H(T)\, T}
	\left( n_\psi n_{\bar\psi} - n_\psi^{\rm eq} n_{\bar\psi}^{\rm eq} \right).
	\label{eq:boltz_sym_yield}
\end{equation}
The annihilation cross section is dominated by the $s$-channel exchange of the heavier scalar $S$. For a Dirac fermion $\psi$ annihilating through a scalar mediator, the process is $p$-wave suppressed in the non-relativistic limit, and the thermally averaged cross section scales as
\begin{equation}
	\langle \sigma v \rangle
	\simeq
	\frac{y_\psi^2 \, \lambda_{SH}^2}{16\pi \, m_S^4} \, T \, m_\psi ,
	\label{eq:sigma_v}
\end{equation}
where the linear dependence on $T$ reflects the $p$-wave nature of the annihilation. The coupling $y_\psi$ entering this expression is chosen such that the annihilation of the symmetric component is efficient at high temperatures, while the late-time decay $S \to \psi\bar\psi$ remains Boltzmann-suppressed.

Integrating the Boltzmann equation for the asymmetry from the onset of production to the present epoch, the final asymmetric yield is obtained as~\cite{Hook:2011}
\begin{equation}
	Y_-^\infty
	=
	\varepsilon
	\int_0^\infty
	\frac{\Gamma_\phi \, n_\phi^{\rm eq}(T)}
	{s(T)\, H(T)\, T}
	\, dT .
	\label{eq:integral_asym}
\end{equation}
By introducing the dimensionless variable $x = m_\phi/T$, the integral can be evaluated analytically. Using~\cite{Gondolo:1991}
\begin{equation}
	\int_0^\infty x^3 K_2(x)\, dx = 8 ,
	\label{eq:bessel_integral}
\end{equation}
the final asymmetry yield simplifies to
\begin{equation}
	Y_-^\infty
	\simeq
	\frac{90}{1.66\, \pi^4}
	\frac{g_\phi}{g_*^{3/2}}
	\frac{M_{\rm Pl}}{m_\phi^2}
	\varepsilon \, \Gamma_\phi ,
	\label{eq:final_yield}
\end{equation}
This result shows that the final asymmetry scales as $Y_-^\infty \propto \varepsilon \, \mu^2 / m_\phi$, where we have used $\Gamma_\phi \propto \mu^2 m_\phi$.

The present-day relic abundance of $\psi$ is determined by the total yield after the symmetric component has been annihilated away~\cite{Petraki:2013,Zurek:2013},
\begin{equation}
	\Omega_{\rm DM} h^2
	=
	2.75 \times 10^8
	\left( \frac{m_\psi}{\rm GeV} \right)
	Y_\Sigma^\infty ,
	\label{eq:relic}
\end{equation}
where $Y_\Sigma^\infty = Y_\psi^\infty + Y_{\bar\psi}^\infty$ is the final total yield. Provided that the annihilation process $\psi\bar\psi \to S \to \text{SM}$ is efficient at high temperatures, the symmetric component is depleted, and one has $Y_\Sigma^\infty \to |Y_-^\infty|$ at late times~\cite{Graesser:2011}, so the relic abundance reduces to
\begin{equation}
	\Omega_{\rm DM} h^2
	=
	2.75 \times 10^8
	\left( \frac{m_\psi}{\rm GeV} \right)
	|Y_-^\infty| .
	\label{eq:relic_asym}
\end{equation}
The final asymmetry yield $Y_-^\infty$ is controlled by the parameters $\mu$, $\varepsilon$, and $m_\phi$, while the total yield $Y_\Sigma^\infty$ is additionally sensitive to the annihilation cross section through $y_\psi$ and $\lambda_{SH}$. The viable parameter space is therefore determined by the interplay between the production of the asymmetry and the efficient removal of the symmetric component, which we explore numerically in the next subsection.

\subsection{The DM abundance}\label{sec3.2}
In this subsection, we present the numerical results obtained from solving the Boltzmann equations governing the production of asymmetric fermionic DM. We analyze the thermal evolution of the asymmetric yield $Y_-$ and investigate its dependence on the relevant model parameters, namely the coupling $\mu$, the CP-violating parameter $\varepsilon$, and the mediator mass $m_\phi$. Throughout this subsection, we fix the DM mass and the auxiliary fermion mass to $m_\psi = 50$~GeV and $m_\chi = 75$~GeV, respectively, while the heavy scalar mass is set to $m_S = 500$~GeV. The production mechanism is dominated by the out-of-equilibrium decay $\phi \to \chi \psi$ (and its CP-conjugate), and the final relic abundance is established once this process becomes inefficient. For each set of parameters, we integrate the Boltzmann equation from $T = 10^4$~GeV down to $T = 10^{-2}$~GeV, ensuring that both the onset and the completion of freeze-in production are fully covered.

\begin{figure}[!htbp]
	\centering
	\includegraphics[width=0.7\textwidth]{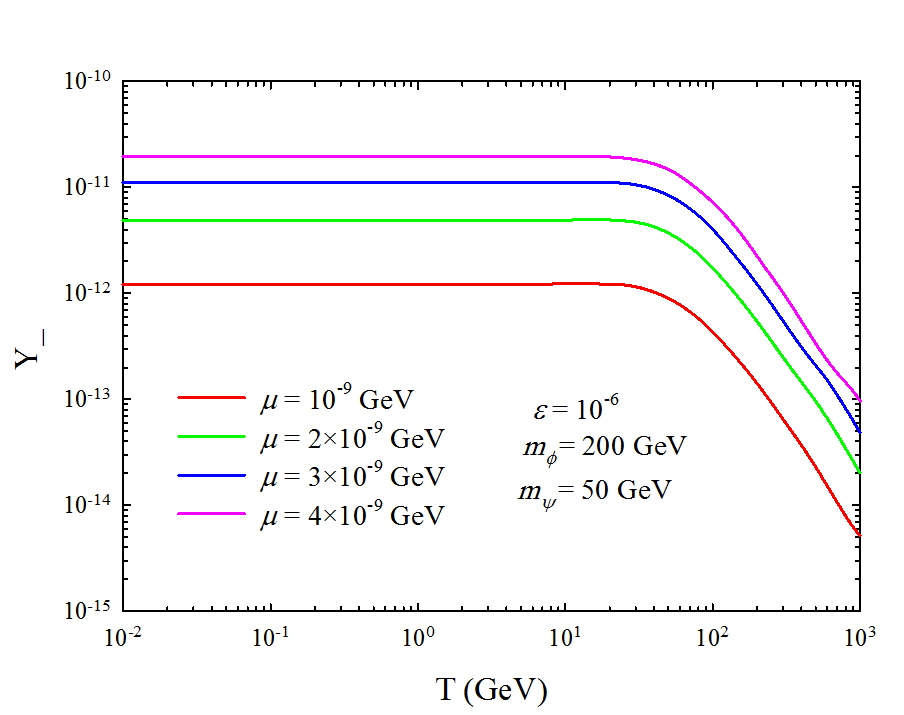}
	\caption{Evolution of the asymmetric yield $Y_-$ as a function of temperature $T$ for different values of the coupling $\mu = 10^{-9}, 2\times10^{-9}, 3\times10^{-9}, 4\times10^{-9}$ GeV, with $\varepsilon = 10^{-6}$, $m_\phi = 200$~GeV, and $m_\psi = 50$~GeV fixed. The temperature decreases from right to left, corresponding to the cooling of the Universe.
	}
	\label{fig:Y_vs_T_mu}
\end{figure}

Figure~\ref{fig:Y_vs_T_mu} shows the evolution of the comoving asymmetric abundance $Y_-$ as a function of temperature $T$ for four different values of the coupling $\mu = 10^{-9}, 2\times10^{-9}, 3\times10^{-9}, 4\times10^{-9}$ GeV, while the CP-violating parameter and the mediator mass are kept fixed at $\varepsilon = 10^{-6}$ and $m_\phi = 200$~GeV. As expected, the production of the DM asymmetry occurs predominantly at temperatures around $T \sim m_\phi$, where the mediator population is still abundant and CP-violating decays efficiently generate the asymmetry. Increasing $\mu$ enhances the decay rate of $\phi$ and leads to a larger final asymmetric yield, with the asymptotic values scaling as $Y_-^\infty \propto \mu^2$, in agreement with the analytical estimate of Eq.~(\ref{eq:final_yield}). In all cases, the yield reaches a constant value at temperatures well below $m_\phi$, confirming that the freeze-in production is completed and the asymmetry is frozen in.

\begin{figure}[!htbp]
	\centering
	\includegraphics[width=0.7\textwidth]{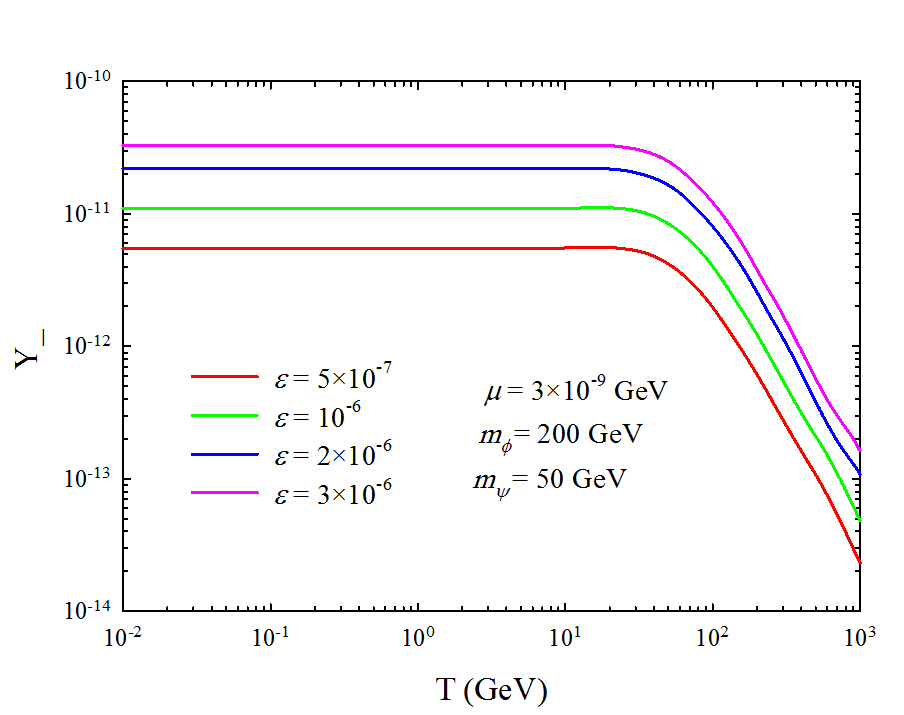}
	\caption{Evolution of the asymmetric yield $Y_-$ as a function of temperature $T$
		for different values of the CP-violating parameter
		$\varepsilon = 5\times10^{-7}, 10^{-6}, 2\times10^{-6}, 3\times10^{-6}$,
		with $\mu = 3\times10^{-9}$~GeV, $m_\phi = 200$~GeV,
		and $m_\psi = 50$~GeV fixed.}
	\label{fig:Y_vs_T_eps}
\end{figure}

Figure~\ref{fig:Y_vs_T_eps} shows the evolution of the asymmetric yield $Y_-$ as a function of temperature for four different values of the CP-violating parameter $\varepsilon = 5\times10^{-7}, 10^{-6}, 2\times10^{-6}, 3\times10^{-6}$, while the coupling and the mediator mass are kept fixed at $\mu = 3\times10^{-9}$~GeV and $m_\phi = 200$~GeV. The final abundance scales approximately linearly with $\varepsilon$, reflecting the fact that the net asymmetry is directly proportional to the CP-violating decay asymmetry of the mediator, as expected from Eq.~(\ref{eq:final_yield}). This behavior confirms that the relic abundance is controlled primarily by the CP-violating source term rather than by thermal equilibrium processes.

\begin{figure}[!htbp]
	\centering
	\includegraphics[width=0.7\textwidth]{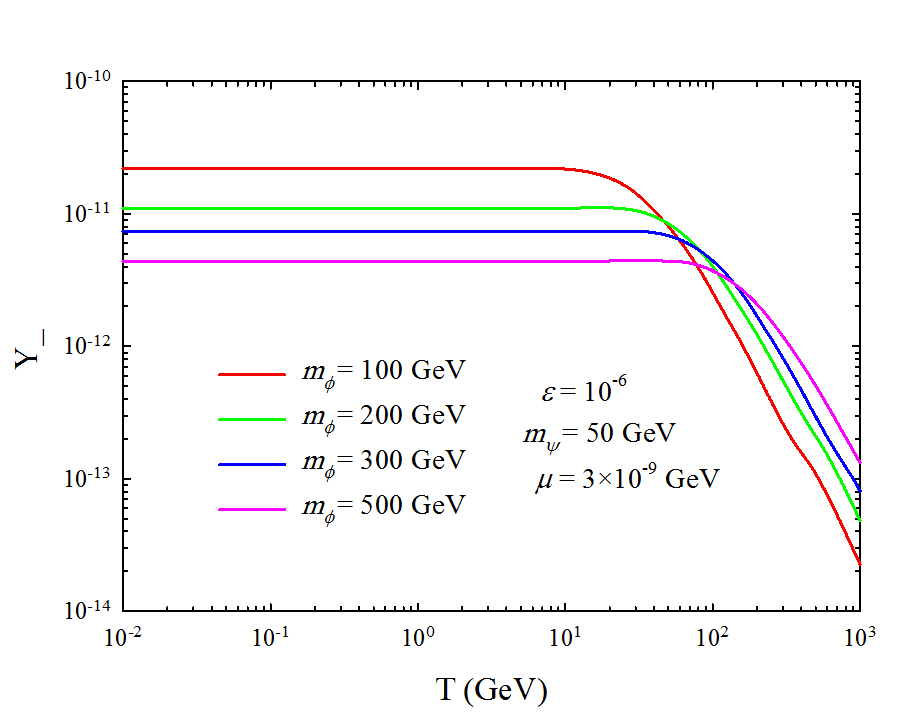}
	\caption{Evolution of the asymmetric yield $Y_-$ as a function of temperature $T$
		for different values of the mediator mass
		$m_\phi = 100, 200, 300, 500$~GeV,
		with $\mu = 3\times10^{-9}$~GeV, $\varepsilon = 10^{-6}$,
		and $m_\psi = 50$~GeV fixed.
	}
	\label{fig:Y_vs_T_mphi}
\end{figure}

Figure~\ref{fig:Y_vs_T_mphi} shows the evolution of the asymmetric yield $Y_-$ as a function of temperature for four different values of the mediator mass $m_\phi = 100, 200, 300, 500$~GeV, while the coupling and the CP-violating parameter are kept fixed at $\mu = 3\times10^{-9}$~GeV and $\varepsilon = 10^{-6}$. Increasing the mediator mass shifts the production epoch to higher temperatures, since the dominant contribution to freeze-in arises at $T \sim m_\phi$, and suppresses the final yield as $Y_-^\infty \propto 1/m_\phi$, in agreement with Eq.~(\ref{eq:final_yield}).

\begin{figure}[!htbp]
	\centering
	\includegraphics[width=0.7\textwidth]{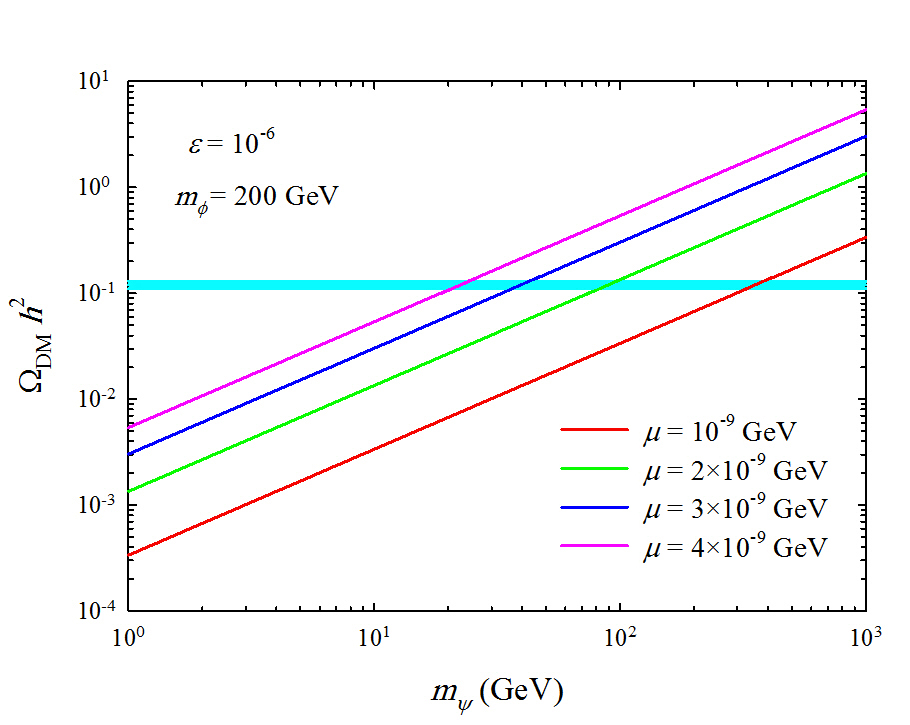}
	\caption{
		Relic density $\Omega_{\rm DM} h^2$ as a function of the DM mass $m_\psi$
		for different values of the coupling
		$\mu = 10^{-9}, 2\times10^{-9}, 3\times10^{-9}, 4\times10^{-9}$~GeV,
		with $\varepsilon = 10^{-6}$ and $m_\phi = 200$~GeV fixed.
		The horizontal shaded band corresponds to the observed DM relic abundance,
		$0.11 \leq \Omega_{\rm DM} h^2 \leq 0.13$,
		centered around the Planck value $\Omega_{\rm DM} h^2 \simeq 0.12$.
	}
	\label{fig:O_vs_mpsi}
\end{figure}

Figure~\ref{fig:O_vs_mpsi} shows the relic density $\Omega_{\rm DM} h^2$ as a function of the DM mass $m_\psi$ for four different values of the coupling $\mu$, while the CP-violating parameter and the mediator mass are kept fixed at $\varepsilon = 10^{-6}$ and $m_\phi = 200$~GeV. In all cases, the relic density exhibits a linear dependence on the DM mass, as expected from the proportionality $\Omega_{\rm DM} h^2 \propto m_\psi \, Y_-^\infty$~\cite{Petraki:2013,Zurek:2013}, since the freeze-in yield is essentially independent of $m_\psi$. The horizontal shaded band corresponds to the observed DM relic abundance, $0.11 \leq \Omega_{\rm DM} h^2 \leq 0.13$, centered around the Planck value $\Omega_{\rm DM} h^2 \simeq 0.12$~\cite{Planck:2020}. The intersection of each diagonal line with this band determines the DM mass range for which the observed relic abundance is reproduced. For $\mu = 3\times10^{-9}$~GeV, the correct relic density is obtained for $m_\psi \approx 40$~GeV, while for the other values of $\mu$ the required DM mass shifts accordingly.

\begin{figure}[!htbp]
	\centering
	\includegraphics[width=0.7\textwidth]{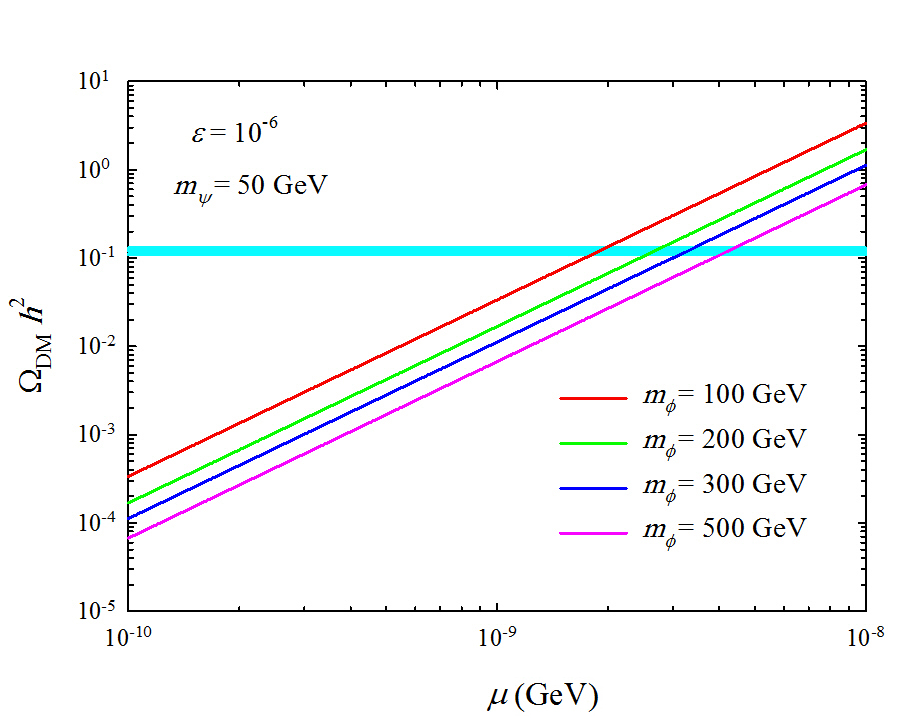}
	\caption{
		Relic density $\Omega_{\rm DM} h^2$ as a function of the coupling $\mu$
		for different values of the mediator mass
		$m_\phi = 100, 200, 300, 500$~GeV,
		with $\varepsilon = 10^{-6}$ and $m_\psi = 50$~GeV fixed.
		The horizontal shaded band has the same meaning as in
		Fig.~\ref{fig:O_vs_mpsi}.
	}
	\label{fig:O_vs_mu}
\end{figure}

Figure~\ref{fig:O_vs_mu} shows the relic density $\Omega_{\rm DM} h^2$ as a function of the coupling $\mu$ for four different values of the mediator mass $m_\phi = 100, 200, 300, 500$~GeV, while the CP-violating parameter and the DM mass are kept fixed at $\varepsilon = 10^{-6}$ and $m_\psi = 50$~GeV. As expected from the analytical estimate $\Omega_{\rm DM} h^2 \propto \mu^2 / m_\phi$, the relic density exhibits a quadratic dependence on the coupling $\mu$, appearing as straight lines with slope~$2$ in the log-log plane. For a fixed value of $\mu$, increasing the mediator mass leads to a smaller relic density, since heavier mediators are less abundantly produced. For the benchmark value $m_\phi = 200$~GeV, the observed relic abundance is reproduced for $\mu \approx 2.7 \times 10^{-9}$, while the other values of $m_\phi$ require correspondingly different couplings.

Finally, we address the fate of the symmetric component of the DM population. In addition to the asymmetric yield $Y_-$, the total yield $Y_\Sigma$ receives contributions from both particles and antiparticles produced in the out-of-equilibrium decays of the mediator. In the early Universe, when the temperature was comparable to the DM mass, the annihilation process $\psi\bar\psi \to S \to \text{SM}$ was kinematically accessible, and the symmetric component of the DM population was depleted. As the Universe cooled below $T \sim m_\psi$, the annihilation rate became Boltzmann-suppressed and the symmetric component effectively disappeared. Since the symmetric component is not protected by any conserved quantum number, only the asymmetric population survives to the present epoch, so that
\begin{equation}
	Y_\Sigma^\infty \simeq |Y_-^\infty| .
	\label{eq:YSigma_limit}
\end{equation}
Under this condition, the present-day relic abundance is determined solely by the asymmetry generated through CP-violating mediator decays, as expressed in Eq.~(\ref{eq:relic_asym}).
   
\section{Phenomenological aspects}\label{sec:aspect}
In this section, we examine the phenomenological implications of the model and confront it with current experimental constraints. Since the DM candidate $\psi$ is a FIMP with extremely feeble couplings to the SM, it is expected to evade most conventional DM searches. Nevertheless, it is essential to verify that the model remains consistent with the strong bounds imposed by direct and indirect detection experiments, collider searches, and cosmological observations. We focus on four main classes of constraints: (i) spin-independent direct detection, (ii) indirect detection through DM annihilation, (iii) collider searches for invisible Higgs decays and missing energy, and (iv) DM self-interactions.

\subsection{Direct detection}\label{sec4.1}
The spin-independent scattering of the DM candidate $\psi$ off a nucleon proceeds through the exchange of the scalar mediator $S$, which couples to the SM via the Higgs portal. The resulting DM-nucleon cross section scales as
\begin{equation}
	\sigma_{\psi N}^{\rm SI}
	\propto
	y_\psi^2 \, \lambda_{SH}^2 ,
	\label{eq:sigma_SI}
\end{equation}
where $y_\psi$ denotes the coupling of $S$ to the dark fermions and $\lambda_{SH}$ is the Higgs-portal coupling. Since both couplings must be sufficiently small to realize the freeze-in regime, the resulting cross section is strongly suppressed. For the parameter space considered in this work, the predicted spin-independent cross section lies far below the current sensitivity of the LUX-ZEPLIN and XENONnT experiments~\cite{LZ:2023,XENONnT:2023}, which constrain spin-independent cross sections down to $\sim 2.2 \times 10^{-48}~\text{cm}^2$ for DM masses around $40$~GeV. Consequently, the model is consistent with all present direct detection bounds, and the predicted signal remains out of reach of current experiments.

\subsection{Indirect detection}\label{sec4.2}
Indirect detection searches for the products of DM annihilation or decay in regions of high DM density, such as dwarf spheroidal galaxies and the Galactic center. The Fermi Large Area Telescope (Fermi-LAT) has placed stringent upper limits on the velocity-averaged annihilation cross section for DM masses between $2$~GeV and $10$~TeV, with typical bounds of order $\langle\sigma v\rangle \lesssim 10^{-25}~\text{cm}^3\,\text{s}^{-1}$ for the $b\bar{b}$ channel~\cite{FermiLAT:2014}. Similar analyses have also been performed for various astrophysical targets using Fermi-LAT data~\cite{Rezaei:2020,Circiello:2024,FermiLAT:2025}. In the present model, however, the annihilation process $\psi\bar\psi \to S \to \text{SM}$ opens only in the high-temperature era, when the symmetric part of the dark population is still abundant and gets efficiently removed. At late times, however, the annihilation rate is strongly Boltzmann-suppressed, since the DM particles are non-relativistic and their number density is too low to produce an observable signal. Consequently, the predicted annihilation cross section lies many orders of magnitude below the Fermi-LAT sensitivity, and the model predicts no observable indirect detection signal. This is a generic feature of freeze-in scenarios, where the DM abundance is produced through rare decays rather than thermal freeze-out, and the late-time annihilation rate is correspondingly negligible.

\subsection{Collider constraints}\label{sec4.3}
The LHC provides a complementary probe of dark sector models through searches for missing transverse energy recoiling against visible SM particles, known as mono-X searches~\cite{ATLAS:2022,CMS:2023}, and through invisible Higgs decays. In the present model, the production of $\psi$ pairs at the LHC is highly suppressed, since the DM candidate couples to the SM only through the Higgs portal with extremely feeble couplings $y_\psi$ and $\lambda_{SH}$. Consequently, the monojet, mono-Higgs, and mono-$Z$ signatures predicted by the model lie far below the current LHC sensitivity. Similarly, the invisible Higgs decay channel $h \to \psi\bar\psi$ is open for $m_\psi = 50$~GeV ($< m_h/2$), but the corresponding partial width is strongly suppressed by the feeble coupling $y_\psi$. For the parameter space considered in this work, the resulting branching ratio lies many orders of magnitude below the current LHC upper bound on invisible Higgs decays~\cite{ATLAS:2022,CMS:2023}. We therefore conclude that the model satisfies the collider constraints considered in this work, and that the predicted signals lie far below the current LHC sensitivity.

\subsection{Self-interacting DM}\label{sec4.4}
Although the DM candidate $\psi$ can in principle self-interact through the exchange of the heavy scalar $S$, the corresponding cross section is extremely suppressed in the FIMP regime. Even for a relatively large coupling $y_\psi \sim 10^{-2}$ and $m_S \sim 500$~GeV, we find $\sigma/m_\psi \sim 10^{-27}$~cm$^2$/g at $m_\psi \sim 40$~GeV, which lies many orders of magnitude below the Bullet cluster bound, $\sigma/m < 0.47$~cm$^2$/g~\cite{Harvey:2015}. This behavior is a generic feature of freeze-in scenarios with fermionic DM: the same feeble couplings that ensure the out-of-equilibrium production also suppress all late-time interactions. In contrast, scalar FIMP DM can naturally accommodate sizable self-interactions, as shown in Refs.~\cite{Zakeri:2026,Zakeri:2018}, where the scalar component satisfies the Bullet cluster constraint while the fermionic one does not. Consequently, the model predicts no observable self-interaction signal.

\section{Concluding remarks}\label{sec:conclud} 
We have explored a minimal and predictive construction in which the observed DM abundance arises from the freeze-in mechanism while carrying a primordial particle–antiparticle asymmetry. The visible sector is supplemented by four new states (a DM Dirac fermion \(\psi\), an auxiliary Dirac fermion \(\chi\), a real scalar \(\phi\) acting as the mediator, and a heavier real scalar \(S\)), all singlet under the SM gauge group and charged under a $Z_{2} \times Z_{2}^{\prime}$ symmetry. This discrete symmetry guarantees the longevity of $\psi$ and eliminates dangerous operators such as $\phi^{3}$ and $\phi H^{\dagger} H$. The dark asymmetry originates from out-of-equilibrium decays of $\phi$, whose CP-violating character is encoded in the parameter $\varepsilon$. The field $S$ enters in two distinct ways: through its loop-level interference with $\phi$, it provides the physical phase needed for a non-vanishing $\varepsilon$, and through its $s$-channel exchange it drives the annihilation of the symmetric part of the dark population. 

We derived and numerically integrated the coupled Boltzmann equations for the asymmetric yield $Y_{-}$ and the total yield $Y_{\Sigma}$. The late-time asymmetry is governed by the combination of $\mu$, $\varepsilon$, and $m_{\phi}$, and our numerical treatment confirms the scaling $Y_{-}^{\infty} \propto \mu^{2} \varepsilon / m_{\phi}$ obtained analytically. We found that the measured value $\Omega_{\mathrm{DM}} h^{2} \simeq 0.12$ emerges naturally over a broad region of parameter space, without any need for thermal contact between the dark and visible sectors. As a representative point, taking $\varepsilon = 10^{-6}$ and $m_{\phi} = 200$~GeV, the observed relic density is reproduced for $\mu \approx 2.7 \times 10^{-9}$~GeV and $m_{\psi} \approx 40$~GeV. We also verified that the symmetric component is efficiently removed through $\psi \bar{\psi} \to S \to \mathrm{SM}$, so that at late times $Y_{\Sigma}^{\infty} \to |Y_{-}^{\infty}|$, and the final abundance is set entirely by the asymmetry.

We then turned to the phenomenological viability of the construction. Spin-independent DM-nucleon scattering proceeds through $S$ exchange and is controlled by the small couplings $y_{\psi}$ and $\lambda_{SH}$; the resulting cross section remains well below the sensitivity of current direct detection experiments such as LUX-ZEPLIN and XENONnT. Indirect searches, in particular those by the Fermi-LAT collaboration, constrain the late-time annihilation rate; because this rate is Boltzmann-suppressed once the Universe cools below the DM mass, the predicted flux is many orders of magnitude smaller than the current limits. Collider probes based on monojet, mono-Higgs, mono-$Z$, and invisible Higgs decays are likewise satisfied: pair production of $\psi$ at the LHC is negligible, and the partial width for $h \to \psi \bar{\psi}$ is suppressed by the feeble coupling $y_{\psi}$, placing the branching ratio far below the LHC bound. Finally, DM self-interactions mediated by $S$ remain extremely weak across the viable parameter space, and the predicted cross section stays many orders of magnitude below the Bullet cluster limit. This is a characteristic feature of freeze-in scenarios: the same suppressed couplings that keep the dark sector out of equilibrium also render all late-time signatures unobservable.

In summary, the framework discussed here offers a simple and internally consistent description of asymmetric fermionic DM produced via freeze-in. Future improvements in direct detection, indirect detection, and collider searches for invisible final states will be able to probe this scenario further.





\end{document}